# Comparative Analysis of Machine Learning and Deep Learning Models for Classifying Squamous Epithelial Cells of the Cervix

Subhasish Das[a], Madhusmita Sethy[c], Prajna Paramita Giri[c], Ashwini K Nanda[b], Satish K Panda[a,b]

[a] *School of Mechanical Sciences, IIT Bhubaneswar, India*

[b] *AI and HPC Research Center, IIT Bhubaneswar, India*

[c] *All India Institute of Medical Sciences, Bhubaneswar*

Corresponding Author: Satish K Panda

First and last name: Satish K, Panda

E-mail address: skpanda@iitbbs.ac.in

**Abstract:**

The cervix is the narrow end of the uterus that connects to the vagina in the female reproductive system. Abnormal cell growth in the squamous epithelial lining of the cervix leads to cervical cancer in females. A Pap smear is a diagnostic procedure used to detect cervical cancer by gently collecting cells from the surface of the cervix with a small brush and analyzing their changes under a microscope. For population-based cervical cancer screening, visual inspection with acetic acid is a cost-effective method with high sensitivity. However, Pap smears are also suitable for mass screening due to their higher specificity. The current Pap smear analysis method is manual, time-consuming, labor-intensive, and prone to human error. Therefore, an artificial intelligence (AI)-based approach for automatic cell classification is needed. In this study, we aimed to classify cells in Pap smear images into five categories: superficial-intermediate, parabasal, koilocytes, dyskeratotic, and metaplastic. Various machine learning (ML) algorithms, including Gradient

Boosting, Random Forest, Support Vector Machine, and k-Nearest Neighbor, as well as deep learning (DL) approaches like ResNet-50, were employed for this classification task.

The ML models demonstrated high classification accuracy; however, ResNet-50 outperformed the others, achieving a classification accuracy of 93.06%. This study highlights the efficiency of DL models for cell-level classification and their potential to aid in the early diagnosis of cervical cancer from Pap smear images.

## Statements and Declarations

### Funding

This study was supported by SERB (SRG/2022/000922) and ICMR (GRANTATHON2/MH-14/2024-NCD-II).

### Competing Interests

The authors have no conflicts of interest to declare.

### Author Contributions

SD (first author): Data preprocessing, network design and training, and manuscript preparation. MS, PPG, AKN, SKP (corresponding author): Fund acquisition, method conceptualization, and manuscript preparation.

### Ethics Approval

No ethical approval was needed as the study was performed on an open-source dataset.

### Data Availability

This study used an open-source dataset: "**Cervical Cancer largest dataset (SipakMed)**" from Kaggle.

## 1. Introduction:

Cervical cancer is the fourth most prevalent cancer among women globally, with over half a million new cases and more than 300,000 deaths reported annually [1]. The primary cause of cervical cancer in most cases is the high-risk subtypes of the human papillomavirus (HPV). Despite being largely preventable, early detection is crucial for saving lives and minimizing treatment costs. A Pap smear, or Papanicolaou smear, involves the microscopic examination of cells collected from the cervix to identify cancerous, pre-cancerous, or other medical conditions. Named after Dr. George N. Papanicolaou, who first described it in 1928, the Pap smear significantly reduces cervical cancer incidence and mortality rates by up to 75% since its introduction [2]. This screening method primarily targets changes in the cervix's transformation zone, often caused by HPV [2]. For population-based cervical cancer screening, visual inspection with acetic acid (VIA) is a cost-effective method with high sensitivity, while Pap smears is preferred for their higher specificity. The current Pap smear analysis method is manual, time-consuming, labor-intensive, and prone to human error. Therefore, an artificial intelligence (AI)-based approach for the automatic identification and classification of cells is needed to reduce the burden on clinicians.

Numerous methods have been proposed in literature to classify the cells present in pap smear images. For example, Bora et al. attempted to classify pap smear images using different machine learning (ML) based classifiers, such as, Least Square Support Vector Machine (LSSVM) and Softmax Regression, and reported good accuracy [22]. Similarly, Liu and coauthors proposed a deep learning (DL) framework, called CVM-Cervix, to perform quick and accurate cervical cell classification from slides [23]. CVM-Cervix first uses a convolutional neural network (CNN) module and then a Visual Transformer module for local and global feature extraction, and subsequently a multilayer perceptron (MLP) module fuses the local and global features for the final classification.

In this study, we aimed to classify the cells present in a Pap smear image into five categories, i.e., superficial-intermediate, parabasal, koilocytes, dyskeratotic, and metaplastic, using different machine learning (ML) and deep learning (DL) algorithms.

## 2. Method:

### 2.1. Image Database

We utilized the open-source "SipakMed Dataset" from Kaggle in our study [3]. This database consisted of 4049 images of isolated cells that were manually cropped from 966 clusters of cell images. These images were acquired through a CCD camera (Infinity 1 Lumenera) adapted to an optical microscope (OLYMPUS BX53F). The dataset consists mainly of five different categories of cells. They are:

- **Superficial-Intermediate cells**: These cells are usually flat with round, oval, or polygonal shapes, with mostly eosinophilic or cyanophilic stained cytoplasm and a central pycnotic nucleus.
- **Parabasal cells**: These are immature squamous cells and are the smallest epithelial cells seen on a typical vaginal smear, with cyanophilic cytoplasm and often a large vesicular nucleus.
- **Koilocytotic cells**: These are pathognomonic cells for HPV infection, and the nucleus of koilocytes usually display various degrees of degeneration, depending on the different stages of infection and the different virus types of infection.
- **Dyskeratotic cells**: These are squamous cells that have undergone premature abnormal keratinization within individual cells or, more often, in three-dimensional clusters and contained orangeophilic cytoplasm.
- **Metaplastic cells**: Essentially, these cells have darker-stained cytoplasm that exhibited greater uniformity of size and shape compared to parabasal cells. Their characteristic is well-defined and have a rounded shape for the cytoplasm.

The distribution of the cells in different classes is outlined in Table 1.

| Category | Number of Images | Number of Cells |
|---|---|---|
| Superficial/Intermediate | 126 | 813 |
| Parabasal | 108 | 787 |
| Koilocytotic | 238 | 825 |
| Metaplastic | 271 | 793 |
| Dyskeratotic | 223 | 813 |
| Total | 966 | 4049 |

Table 1. The types of cells and their count in SIPaKMeD Database

Fig. 1 shows a sample image from each class of the data set.

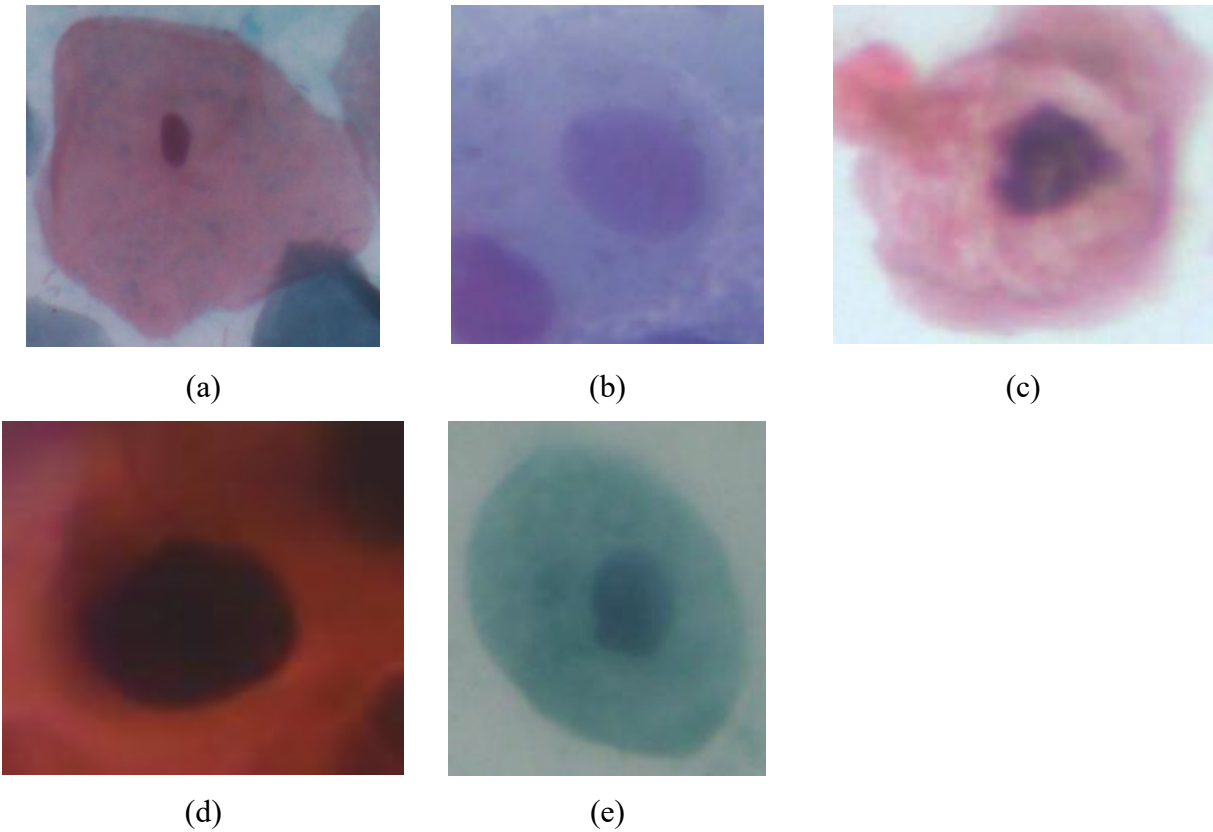

Figure 1: (a) Superficial-Intermediate cells (b) Parabasal cells (c) Koilocytotic cells (d) Dyskeratotic cells (e) Metaplastic cells

### 2.2. ML and DL methods for classification

Image classification is a critical task in computer vision that involved categorizing images into predefined labels or classes. Traditional ML techniques, such as Support Vector Machines (SVM) and k-Nearest Neighbors (k-NN), are often based on handcrafted features and classical algorithms. In contrast, DL models automatically learn hierarchical features directly from the raw image, which help to improve their performance significantly.

*Machine Learning Models*

We tested four different ML approaches for the classification task. The features from the images were extracted using the Histogram of Oriented Gradient (HOG) algorithm. The basic hypothesis of the algorithm is that the local object appearance and shape can be characterized rather well by the distribution of local intensity gradients or edge directions, even without precise knowledge of the corresponding gradient or edge positions. The aim of HOG is to describe an image with a local-oriented gradient histogram. These histograms represents the occurrences of specific gradient orientations in local parts of the image. HOG feature extraction consists of several important steps as outlined in the work of Putra et al. [21]. We extracted different features from Pap smear images using the HOG tool and used them as inputs to our ML models. The workflow for ML model training is shown in Fig. 2.

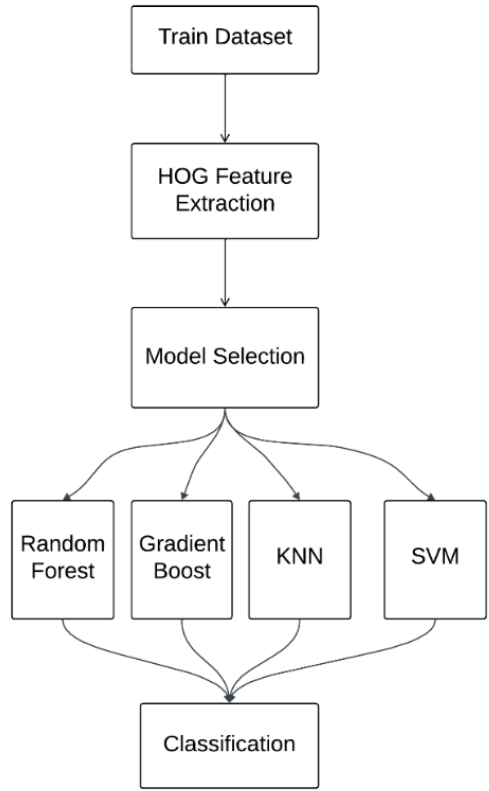

Figure 2: Workflow for the classification of cervical cell subtypes using feature extraction and model selection

**Random Forest**

Random forests, introduced by Breiman, uses randomization to build multiple decision trees and aggregates their outputs for classification (via voting) or regression (via averaging). Randomization occurres in two ways: (1) Bootstrap sampling creates training subsets by sampling with replacement, leaving about 37% of the data as "out-of-bag" samples for model testing. (2) At each decision node, a random subset of predictors (typically $\sqrt{p}$ for p predictors) is tested, selecting the split that best separated the data. This process continues until a stopping criterion is met. Hundreds of such trees are grown to form the forest, ensuring robust and accurate predictions [4].

**Gradient Boosting**

Gradient boosting machines (GBMs) iteratively fit new models to correct errors made by previous ones, aligning with the negative gradient of the ensemble's loss function. While the loss function can vary, common choices included squared-error loss for regression tasks, resulting in sequential error correction. GBMs offer flexibility, allowing researchers to customize or design task-specific loss functions, though this often requires trial and error. Despite their experimental nature, GBMs are straightforward to implement and achieve notable success in practical applications and machine-learning competitions [5].

**k-Nearest Neighbors (KNN)**

The k-Nearest Neighbors (kNN) is a simple, non-parametric classification method. It classifies a data point based on majority voting or distance-weighted voting among its k nearest neighbors. The choice of k significantly impacts the performance, with the optimal value is often determined through experimentation by testing multiple k values and selecting the one with the best results [6].

**Support Vector Machines (SVM)**

Support Vector Machine (SVM), introduced by Boser, Guyon, and Vapnik in 1992, is a supervised learning method for classification and regression. It uses a hypothetical space of linear functions in a high-dimensional feature space, guided by the Structural Risk Minimization principle to enhance generalization. SVM gained recognition for achieving high accuracy in tasks like handwriting recognition and have been widely applied in pattern classification and regression. Its ability to maximize predictive accuracy while avoiding overfitting make it a powerful tool in machine learning [7].

**Deep Learning Model**

**ResNet-50**

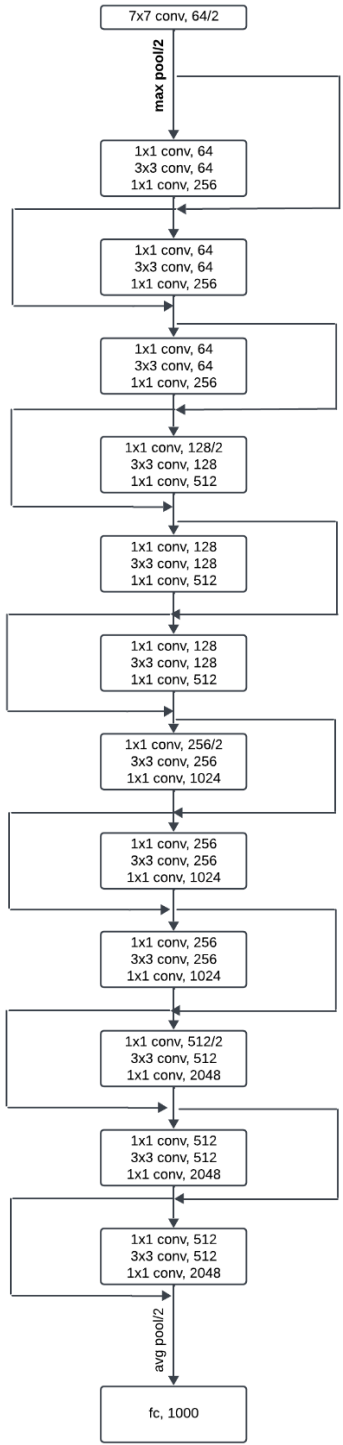

Fig. 3: Resnet-50 Architecture

ResNet-50 is a convolutional neural network that belongs to the Residual Network (ResNet) family. Its primary purpose is to overcome the challenges of training very deep networks and it specifically deals with the vanishing gradient problem. It consists of 50 layers that are structured in a series of residual blocks. It uses skip (shortcut) connections to bypass one or more layers. This enabled the network to learn residual functions, which effectively simplifies the optimization process and allows for deeper architecture without performance degradation. The design of ResNet-50 includes bottleneck layers within its residual blocks, which use 1x1 convolutions to reduce and then restore dimensions with a 3x3 convolution, optimizing computational efficiency (see Fig. 3 for network architecture). With its robust and scalable design, ResNet-50 has shown promising results on various tasks, such as image classification and object detection, and has widely been used in research [8].

We modified the final layer of the network with a new linear layer matching the number of classes in the dataset. The model used ImageNet pre-trained weights for feature extraction and was trained over 500 epochs with a batch size of 32 for both training and evaluation. The Adam optimizer was employed with a learning rate of 0.001.

To analyze the performance of the model, we used the confusion matrix and the one-vs-rest multiclass Receiver Operating Characteristic (ROC) curve. A confusion matrix summarizes prediction outcomes (True Positives, True Negatives, False Positives, False Negatives) for all classes, offering insights into misclassifications and class-wise accuracy, whereas the one-vs-rest approach computes a multi-class ROC curve for each class by treating it as positive and the others as negative, visualizing classifier performance through true positive and false positive trade-offs for each class.

We used operations, such as resizing, center cropping, tensor conversion, normalization, flipping, noise addition, and contrast change, to augment our dataset. We split the dataset into three categories, i.e., train, validate, and test sets, in the proportion of 8:1:1 to execute supervised learning on them. Finally, we evaluated the performance of the models and compared them.

## 3. Results

We observed that the ML and DL models were able to classify the cells with high accuracy. Among the ML models, the SVM was able to achieve the highest accuracy of 70.64%. The ResNet-50 model, however, outperformed all ML models with an accuracy of 93.06%. Especially for Parabasal and Superficial-Intermediate cells, it showed very high accuracy of 98.89% and 95.83%, respectively. The performance of the ML models on the test set are outlined in the Table 2.

| Sl. No. | Algorithm Applied | Accuracy |
|---|---|---|
| 1 | Random Forest | 50.992% |
| 2 | Scikit-learn Gradient Boost | 57.143% |
| 2 | XGBoost | 57.341% |
| 4 | CatBoost | 47.024% |
| 5 | kNN | 58.333% |
| 6 | SVM | 70.635% |
| 7 | ResNet-50 | 93.056% |

Table 2. Performance of ML and DL models

Although computationally expensive, ResNet-50 was able to provide very promising results. The observed classification accuracies for each cell subtype were: 85.98% for Koilocyte cells, 93.27% for Dyskeratotic cells, 92.52% for Metaplastic cells, 98.89% for Parabasal cells, and 95.83% for Superficial-Intermediate cells. Fig. 4 illustrate the confusion matrix and the ROC curve for the classification task.

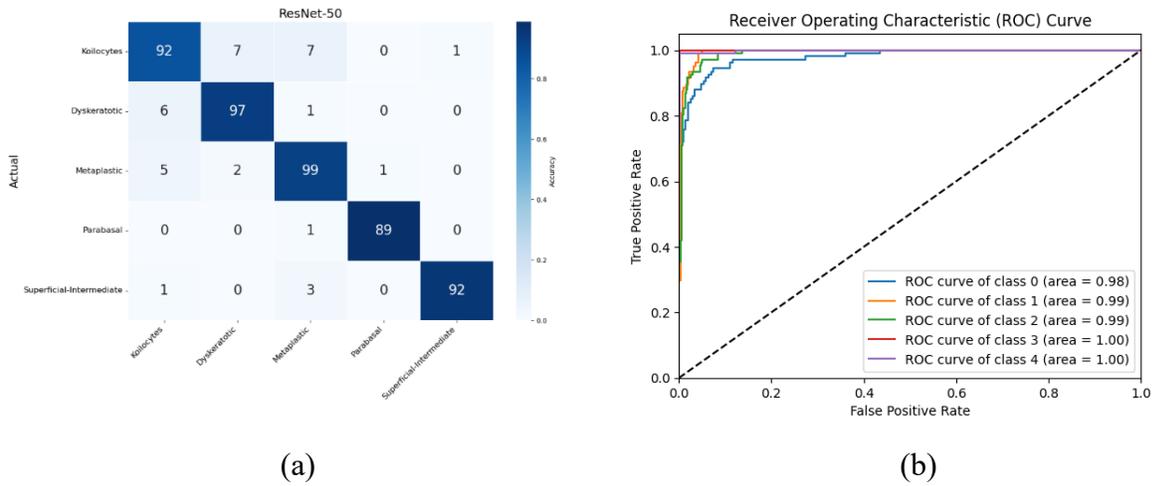

Figure 4: a) Confusion matrix for the classification task using ResNet-50 b) Multi-Class ROC Curve: Performance evaluation for five cervical cell subtypes—Class 0: Koilocytes, Class 1: Dyskeratotic, Class 2: Metaplastic, Class 3: Parabasal, and Class 4: Superficial-Intermediate

## 4. Discussions

In this study, we deployed numerous ML and DL models for the classification of squamous epithelial cells of the cervix into five subcategories. The ML models showed promising results for classification, but the ResNet-50 model outperformed the others with an accuracy of 93.06%. The training process for ResNet-50 was found to be computationally expensive. On the other hand, we were able to train ML algorithms with minimal resources but they did not provide good accuracy. For clinical acceptance, accuracy plays a vital role, and thus, we need to rely on deep learning networks rather than simple ML-based approaches.

The use of HOG for feature extraction was effective for feature extraction and classifying Pap smear images because it captured essential structural details of the cells. It excelled at detecting edge directions and shape descriptors, making it suitable for identifying cell boundaries and distinguishing normal cells from abnormal ones. This was particularly important in Pap smear analysis, where irregularities in the shapes and sizes of nuclei or cytoplasm can indicate potential abnormalities.

In addition, HOG captured texture variations, which were crucial for distinguishing between normal cells with uniform textures and abnormal cells with disrupted patterns. Its robustness to changes in lighting or staining ensured reliable feature extraction even with varying image quality. By producing a compact feature vector, HOG efficiently highlighted meaningful information, making it ideal for machine learning models to accurately classify normal and abnormal cells in cervical cytology.

We observed that our ML models failed to achieve good accuracy as compared to that of the DL model. The DL model, though computationally expensive, achieved an accuracy of 93.06%. The major drawbacks of ML models are their high dependency on feature engineering and their frequent reliance on simpler architectures, which may not be capable of capturing highly non-linear relationships in data. We believe this work will assist medical practitioners for effective diagnosis of cervical cancer at an earlier stage and will help to reduce the mortality rate worldwide.

# REFERENCES

1. Cohen, P. A., Jhingran, A., Oaknin, A., & Denny, L. (2019). Cervical cancer. The Lancet, 393(10167), 169-182.

2. Mehta, V., Vasanth, V., & Balachandran, C. (2009). Pap smear. Indian journal of dermatology, venereology and leprology, 75, 214.

3. Plissiti, M. E., Dimitrakopoulos, P., Sfikas, G., Nikou, C., Krikoni, O., & Charchanti, A. (2018, October). Sipakmed: A new dataset for feature and image based classification of normal and pathological cervical cells in pap smear images. In 2018 25th IEEE international conference on image processing (ICIP) (pp. 3144-3148). IEEE.

4. Rigatti, S. J. (2017). Random forest. *Journal of Insurance Medicine*, *47*(1), 31-39.

5. Natekin, A., & Knoll, A. (2013). Gradient boosting machines, a tutorial. *Frontiers in neurorobotics*, *7*, 21.

6. Guo, G., Wang, H., Bell, D., Bi, Y., & Greer, K. (2003). KNN model-based approach in classification. In *On The Move to Meaningful Internet Systems 2003: CoopIS, DOA, and ODBASE: OTM Confederated International Conferences, CoopIS, DOA, and ODBASE 2003, Catania, Sicily, Italy, November 3-7, 2003. Proceedings* (pp. 986-996). Springer Berlin Heidelberg.

7. Jakkula, V. (2006). Tutorial on support vector machine (svm). *School of EECS, Washington State University*, *37*(2.5), 3.

8. Koonce, B., & Koonce, B. E. (2021). *Convolutional neural networks with swift for tensorflow: Image recognition and dataset categorization* (pp. 109-123). New York, NY, USA: Apress.

9. Shingleton, H. M., Patrick, R. L., Johnston, W. W., & Smith, R. A. (1995). The current status of the Papanicolaou smear. *CA: a cancer journal for clinicians*, *45*(5), 305-320.

10. Affonso, C., Rossi, A. L. D., Vieira, F. H. A., & de Leon Ferreira, A. C. P. (2017). Deep learning for biological image classification. *Expert systems with applications*, *85*, 114-122.

11. Abuzaid, N. N., & Abuhammad, H. Z. (2022). Image SPAM Detection Using ML and DL Techniques. *International Journal of Advances in Soft Computing & Its Applications*, *14*(1).

12. LeCun, Y., Bengio, Y., & Hinton, G. (2015). Deep learning. *nature*, *521*(7553), 436-444.

13. Krizhevsky, A., Sutskever, I., & Hinton, G. E. (2012). Imagenet classification with deep convolutional neural networks. *Advances in neural information processing systems*, *25*.

14. Rawat, W., & Wang, Z. (2017). Deep convolutional neural networks for image classification: A comprehensive review. *Neural computation*, *29*(9), 2352-2449.
15. Jain, A. K., Duin, R. P. W., & Mao, J. (2000). Statistical pattern recognition: A review. *IEEE Transactions on pattern analysis and machine intelligence*, *22*(1), 4-37.
16. Chen, T., & Guestrin, C. (2016, August). Xgboost: A scalable tree boosting system. In *Proceedings of the 22nd acm sigkdd international conference on knowledge discovery and data mining* (pp. 785-794).
17. Prokhorenkova, L., Gusev, G., Vorobev, A., Dorogush, A. V., & Gulin, A. (2018). CatBoost: unbiased boosting with categorical features. *Advances in neural information processing systems*, *31*.
18. Fawcett, T. (2006). An introduction to ROC analysis. *Pattern recognition letters*, *27*(8), 861-874.
19. Hand, D. J., & Till, R. J. (2001). A simple generalisation of the area under the ROC curve for multiple class classification problems. *Machine learning*, *45*, 171-186.
20. Sokolova, M., & Lapalme, G. (2009). A systematic analysis of performance measures for classification tasks. *Information processing & management*, *45*(4), 427-437.
21. Putra, F. A. I. A., Utaminingrum, F., & Mahmudy, W. F. (2020). HOG feature extraction and KNN classification for detecting vehicle in the highway. *IJCCS (Indonesian Journal of Computing and Cybernetics Systems)*, *14*(3), 231-242.
22. Bora, K., Chowdhury, M., Mahanta, L. B., Kundu, M. K., & Das, A. K. (2016, December). Pap smear image classification using convolutional neural network. In *Proceedings of the tenth Indian conference on computer vision, graphics and image processing* (pp. 1-8).
23. Liu, W., Li, C., Xu, N., Jiang, T., Rahaman, M. M., Sun, H., ... & Grzegorzek, M. (2022). CVM-Cervix: A hybrid cervical Pap-smear image classification framework using CNN, visual transformer and multilayer perceptron. *Pattern Recognition*, *130*, 108829.